\documentclass[aps,prl,twocolumn,showpacs]{revtex4} 
\newcommand\beq{\begin{equation}}
\newcommand\eeq{\end{equation}}
\newcommand\bea{\begin{eqnarray}}
\newcommand\eea{\end{eqnarray}}
\def\half{{1 \over 2}}

\usepackage{graphics}

\begin{document}
\title{Shape invariant potentials in SUSY quantum mechanics 
and periodic orbit theory}

\author{Rajat K. Bhaduri}
\author{Jamal Sakhr} 
\author{D.W.L. Sprung} 
\affiliation{Department of Physics and Astronomy, McMaster University,
 Hamilton, Ontario L8S 4M1, Canada}
\author{Ranabir Dutt}     
\affiliation{Physics Department, Visva Bharati University, Santiniketan 73125, India}
\author{Akira Suzuki}
\affiliation{Department of Physics, Tokyo 
University of Science, Tokyo 162-8601, Japan }

\date{\today} 

\begin{abstract}
We examine shape invariant potentials (excluding those that are 
obtained by scaling) in supersymmetric quantum mechanics from the 
stand-point of periodic orbit theory. An exact trace formula for 
the quantum spectra of such potentials is derived. Based on this 
result, and Barclay's functional relationship for such potentials,  
we present a new derivation of the result that  the lowest order SWKB 
quantisation rule is exact. 
\end{abstract} 


\maketitle 

In non-relativistic quantum mechanics certain potentials are 
amenable to exact analytic solution. 
For a subset of these soluble potentials, the energy spectrum 
may be expressed explicitly as an algebraic function of 
a single quantum number. Such potentials occur either in one space 
dimension, or are central potentials in higher dimensions. For the 
latter, an effective potential in the radial variable can be defined 
for each partial wave. Some examples of such potentials are 
Coulomb, harmonic oscillator, Morse, Rosen-Morse, etc.~\cite{cooper1}. 
These potentials also have the property that the lowest order WKB 
quantisation rule, together with the appropriate Maslov index (that may 
change from potential to potential~\cite{migdal}) , leads to 
exact results. For central potentials, the Langer 
prescription~\cite{langer} for the centrifugal barrier, together with 
half-integer quantisation, can also be employed~\cite{barclay}. 

In supersymmetric (SUSY) quantum mechanics, these exactly solvable 
potentials are found to be translationally shape invariant~\cite{gend}. 
Combining SUSY and WKB, Comtet et al.~\cite{comtet} found that the 
lowest order SWKB  calculation needs neither the Maslov index nor the 
Langer correction to yield the exact result. The purpose of the 
present paper is to understand this result from the point of view of 
the periodic orbit theory (POT)~\cite{brack}, rather than the higher 
order WKB corrections~\cite{khare}. Regarding the latter, we should 
point out a 
largely overlooked paper by  Barclay~\cite{barclay}, in which he 
showed that the higher order WKB terms converge in these potentials 
to yield an energy-independent correction. The  latter may be 
absorbed in the Maslov index of the lowest order term. For SWKB, all 
the higher order terms vanish. Although we do not make the WKB 
expansion, we arrive at the same result in a novel application of 
POT.   

We first set the notation by reviewing the relevant equations of SUSY QM. 
Consider a potential $V(x;a_1)$ of a single variable $x$, and a set of 
parameters denoted by $a_1$. One defines a ``super potential'' 
$$
W(x;a_1)=-{\hbar\over {\sqrt{2m}}}{\phi^{\prime}_0(x)\over {\phi_0(x)}}~,
$$ 
where $\phi_0(x)$ is the ground state solution of the Schr\"odinger 
equation at energy $E_0$ for the potential $V(x,a_1)$, and a prime 
denotes the spatial derivative. Let us define 
 \begin{equation}                      
V_1(x;a_1)=(V(x;a_1)-E_0)~, 
\end{equation}
so that the ground state energy of the Hamiltonian 
$$
H_1=-{\hbar^2\over {2m}}{d^2\over {dx^2}}+V_1(x;a_1)
$$
lies at zero energy, i.e., $E_0^{(1)}=0$. Then it is easy to show that 
$$
V_1(x;a_1)=W^2(x;a_1)-{\hbar\over {\sqrt{2m}}}W^{\prime}(x;a_1)~.
$$
The SUSY partner Hamiltonian $H_2$ has the potential $V_2(x;a_1)$, and  
has an energy spectrum identical to that of $H_1$, except for the 
absence of the zero-energy state. The ground state of $H_2$, denoted 
by $E_0^{(2)}$ coincides with the first excited state $E_1^{(1)}$ of 
$H_1$, and so on. The partner potential $V_2(x;a_1)$ is 
$$
V_2(x;a_1)=W^2(x;a_1)+{\hbar\over {\sqrt{2m}}}W^{\prime}(x;a_1)~.
$$
Shape invariance in the partner potentials is defined by the relation 
 \begin{equation}                       
V_2(x;a_1)=V_1(x;a_2)+R(a_1)~,
\label{shape}
\end{equation}
where the new parameters 
$a_2$ are some function of $a_1$, and the remainder $R(a_1)$ is 
independent of the variable $x$. We restrict our consideration of 
shape invariance to those cases where $a_2$ and $a_1$ are related by 
translation, $a_2=a_1+\alpha$. It is then straight forward to show, 
by constructing a hierarchy of Hamiltonians that 
the complete eigenvalue spectrum of $H_1$ is given by~\cite{cooper1} 
 \begin{eqnarray}                       
E_n^{(1)}&=&\sum_{k=1}^n R(a_k)~,~~n\geq 1~,\\
E_0^{(1)}&=&0~.
\end{eqnarray}
The RHS of the above may be expressed as a monotonic function $f_1(n)$ of the 
quantum number $n$, so that 
 \begin{equation}                       
E_n^{(1)}=f_1(n)~;~~~~f_1(0)=0~.
\label{mark}
\end{equation} 
For the shape invariant potentials we consider here, $f_1(n)$ is an 
algebraic function. Using this property, we proceed to obtain an 
exact expression for the quantum density of states of $H_1$ in the 
spirit of periodic orbit theory. This entails a division of the 
density of states into a smooth and an oscillating part as a function 
of a continuous classical variable $E$. To this end, we may write  
 \begin{equation}                       
\delta (E-E_n^{(1)})=\delta (E-f_1(n))=\delta (n-F_1(E)) F_1^{\prime}(E), 
\label{mark4}
\end{equation}
where the relation $E=f_1(n)$ has been inverted to define 
 \begin{equation}                       
n=F_1(E)~.
\label{mark1}
\end{equation}   
(When the mapping between $n$ and $E_n$ is $(1,1)$ there is no 
difficulty in the inversion. But to define the derivative 
$F_1^{\prime}(E)$ requires an extension from the discrete to the 
continuum, and that is not unique. In the case of shape invariant 
potentials we have an algebraic relation which solves that problem.)
For the spectrum under consideration, $f_1(0)=0$ implies the condition 
 \begin{equation}                       
F_1(0)=0~.
\label{mark2}
\end{equation} 
The quantum density of states $g_1(E)$ for the discrete spectrum of 
$H_1$ is defined as 
 \begin{equation}                       
g_1(E)=\sum_{n=0}^{\infty}d(n) \delta(E-E_n^{(1)})~,
\end{equation}
where $d(n)$ is the degeneracy of states at $E=E_n$. Writing 
$d(n)=d(F_1(E))= D(E)$, and using Eq. (\ref{mark4}), we obtain 
 \begin{equation}                       
g_1(E)=D(E) F_1^{\prime}(E)\sum_{n=0}^{\infty} \delta(n-F_1(E))~.
\end{equation} 
($D(E)=1$ for one-dimensional potentials). We now use the identity 
 \begin{equation}                       
\sum_{n=0}^{\infty} \delta(n-x)=\sum_{k=-\infty}^{\infty} e^{2i\pi kx}~,~~
 x\geq 0~,
\end{equation}
to obtain the desired expression~\cite{brack,jennings}
 \begin{equation}                       
g_1(E)=D(E) F_1^{\prime}(E)\left[1+2\sum_{k=1}^{\infty} \cos[2\pi k F_1(E)]
\right]~.
\label{first}
\end{equation}
For a given $F_1(E)$, this is an {\it exact} expression for the quantum 
density of states $g_1(E)$. It is in the form of a trace formula 
in POT~\cite{brack,Balian}  when $F_1(E)$ (to within a dimensionless 
additive constant $\eta$) is related to  the action $S_1(E)$ of the primitive 
classical periodic orbit of the potential $V_1(x)$ :   
 \begin{eqnarray}                       
{S_1(E)\over h}&=&F_1(E)+\eta~, \label{sec} \\  
S_1(E)&=& 2 \sqrt{2m}\int_{x_1}^{x_2}\sqrt{E-V_1}~dx~.
\label{constant}
\end{eqnarray}
In the above, $x_1$ and $x_2$ are the classical turning points at which
$E=V_1(x)$ (for simplicity in notation, we write $V_1(x,a_1)=V_1(x)$). 
The ($h-$independent constant) $\eta$ may be determined by using 
Eq. (\ref{sec}), and 
applying the condition given by Eq. (\ref{mark2}) for $E=0$. We then 
obtain  
 \begin{equation}                       
\eta = {S_1(0)\over h}~.
\label{maslov}
\end{equation}
We may prove Eq. (\ref{sec}) by noting that the (smooth) 
Thomas-Fermi density of states, given by the first term on the RHS of 
Eq. (\ref{first}), is the Laplace inverse of the classical canonical 
partition function~\cite{ross} of the Hamiltonian 
$H_1^{cl}(x,p)=p^2/2m+V_1(x)$ : 
 \beq                                   
F_1^{\prime}(E)={\cal{L}}^{-1}_E~ Z_1^{cl}(\beta)~~
               ={1\over {2\pi i}}\int_{c-i\infty}^{c+i\infty}Z_1^{cl}
(\beta) e^{\beta E}~d\beta.
\label{jim}
\eeq
Since 
 \bea                                   
Z_1^{cl}(\beta)&=&{1\over h}\int \exp[-\beta\,H_1^{cl}(x,p)]~dx dp \nonumber \\
               &=&{1\over {2\pi\hbar}}\sqrt{2m\pi\over \beta}
                  \int_{-\infty}^{\infty} \exp[-\beta V_1(x)] dx~,
\eea
it follows from (\ref{jim}) that 
 \beq                                   
F_1^{\prime}(E)={\sqrt{2m}\over {2\pi \hbar}} 
\int_{x_1}^{x_2} {dx\over {\sqrt{[E-V_1(x)]}}}~.
\eeq
From this, Eq. (\ref{sec}) follows on integration over energy. Using 
Eq. (\ref{mark1})  together with (\ref{sec}, \ref{constant}), we 
obtain the important result that the lowest order WKB quantisation 
rule is {\it exact} for $V_1$ : 
 \beq                                   
S_1(E)=\oint p(x) dx= (n+\eta)h~,
\label{tuk}
\eeq
where $p(x)=\sqrt{2m[E-V_1(x)]}$. We also see that the constant $\eta$ is 
the so called Maslov index which may vary from one potential to the 
other.

The Maslov index $\eta$ may be eliminated from the quantisation rule by 
employing the superpotential formalism, and the result of Barclay and  
Maxwell~\cite{barcmax}. They made the important observation that the shape 
invariant class of potentials under consideration obey one or 
other of the following equations: \\ 
Class I 
 \beq                                   
\frac{\hbar}{\sqrt{2m}}
{dW\over dx}=A+B W^2(x)+C W(x)~,  
\label{class1}
\eeq
or 
Class II
 \beq                                    
\frac{\hbar}{\sqrt{2m}}
{dW\over dx}=A+B W^2(x)+C W(x) \sqrt{(A+B W^2)}~,
\label{class2}
\eeq
where A, B and C are constants. Using these equations, we now show 
that $S_1(E)$, as defined by Eq. (\ref{constant}), obeys the relation 
($x_{1s}$, $x_{2s}$ are the turning points in SWKB) 
\beq                                   
S_1(E)= 2 \sqrt{2m}\int_{x_{1s}}^{x_{2s}}\sqrt{E-W^2}~dx~+ h \eta~.
\label{semi}
\eeq
To this end, note that the action $S_1$ can be expressed as an 
inverse Laplace transform 
 \beq                                   
S_1(E)=\sqrt{2m\pi}~ {\cal L}^{-1}_E \int_{-\infty}^{\infty}
{e^{-\beta[W^2- \hbar
 W^{\prime}/\sqrt{2m}]}\over {\beta^{3/2}}} dx~. 
\eeq
At this point, for simplicity of notation, let us 
temporarily put $\hbar/\sqrt{2m}=\gamma $. 
Expanding the exponential in powers of  $W^{\prime}$, we have 
 \bea                                   
&& S_1(E) = \nonumber \\ 
&& \sqrt{2m\pi} {\cal L}^{-1}_E \int_{-\infty}^{\infty}{e^{-\beta W^2}
\over {\beta^{3/2}}}\left(1+\sum_{k=0}^{\infty} {(\gamma \beta 
W^{\prime})^{k+1}\over {(k+1)!}}\right)~ dx~.~\\ 
      &=&2\sqrt{2m}\int_{x_{1s}}^{x_{2s}}\sqrt{E-W^2} dx + 
\nonumber \\ && \quad 
\sum_{k=0}^{\infty} {\hbar\over (k+1)!} {\partial^k\over\partial E^k} 
\int_{-\sqrt{E}}^{\sqrt{E}}{(\gamma W^{\prime})^k\over {\sqrt{E-W^2}}} dW~.
\label{kyabat}
\eea
Note that now the limits in $x$ are replaced by the condition 
$W^2(x)=E$. The integral for $k=0$ may be done immediately, yielding 
$\pi$. To evaluate the integrals for integer $k\geq 1$, we assume 
that $\gamma\, W^{\prime}$ obeys Barclay's equation (\ref{class1}) 
(class I) or (\ref{class2}) (class II). 

For class I, we require integrals of the type 
 \beq                                   
I_k=\int_{-\sqrt{E}}^{\sqrt{E}} {(A+B W^2+C W)^k\over \sqrt{E-W^2}} dW~.
\eeq   
On expanding the numerator, terms with odd powers of $W$ vanish on 
integration. One now sees that only the piece of $I_k$ involving the 
highest power of $W^2$ survives the differentiation in Eq. (\ref{kyabat}). 
Consider the integral with $W^{2k}$. With the substitution $W = 
\sqrt{E} \sin \theta$ 
 \bea                                   
\int_{-\sqrt{E}}^{\sqrt{E}} { W^{2k}\over \sqrt{E-W^2}} dW &=&E^k
\int_{-\pi/2}^{\pi/2} \rm{sin}^{2k}\theta d\theta~ \label{rep}\\
&=&E^k {(2k-1)!!\over {(2k)!!}}\pi~. 
\eea   
Accordingly, Eq. (\ref{kyabat}) reduces to    
 \bea                                   
S_1(E) &=& 2\sqrt{2m}\int_{x_{1s}}^{x_{2s}}\sqrt{E-W^2} dx \,\, + \nonumber \\ 
&& \quad \hbar \pi\left(1+ \sum_{k=1}^{\infty} {B^k (2k-1)!!\over 
{(k+1)(2k)!!}}\right)~.
\eea
By construction, $W^2(x)$ has coincident turning points at $E=0$, so the 
first term on the RHS above vanishes at this energy.
Comparing with Eq. (\ref{maslov}), we deduce that 
 \bea                                   
\eta &=& {1\over 2}\left(1+ \sum_{k=1}^{\infty} 
{B^k (2k-1)!!\over {(k+1)(2k)!!}}\right) \nonumber \\ 
&=& \frac{1}{B} [ 1 - \sqrt{1-B}]~.
\label{eta}
\eea
Note, from Eq. (\ref{class1}), that $B$ is independent of 
Planck's constant $h$. Comparing now with Eq. (\ref{sec}), we deduce 
our main result 
 \beq                                   
2\pi\hbar F_1(E)=\sqrt{2m}\oint~\sqrt{E-W^2} dx~. 
\label{fine}
\eeq
Using Eq. (\ref{mark1}) we get as  
the {\it exact} result the SWKB expression 
 \beq                                   
\oint~\sqrt{2m(E-W^2)} dx~=2\pi\hbar n~,~~ n=0,1,2,3, \cdots 
\label{cos}
\eeq
which yields the quantum spectrum of $V_1(x)$.

A similar derivation may be carried through for class II superpotentials 
obeying Eq. (\ref{class2}). The starting point, as before, is 
Eq. (\ref{kyabat}), and the integral to be considered is now of the form 
 \beq                                   
J_k=\int_{-\sqrt{E}}^{\sqrt{E}} {(A+BW^2)^k\left(1+{CW\over 
{\sqrt{A+BW^2}}}\right)^k \over {\sqrt{E-W^2}}}~dW~. 
 \eeq 
The second bracketed term in the numerator on the RHS may be expanded 
binomially, and the odd-powered terms in $W$ vanish on integration. We 
then have $J_k = $ 
 \beq                                   
\sum_{n=0}^{n_{max}} {k!\over {(k-2n)!(2n)!}}\int_{-\sqrt{E}}^{\sqrt{E}}
{(A+BW^2)^{k-n}(CW)^{2n}\over {\sqrt{E-W^2}}} dW~,
\eeq
where $n_{max}=k/2$ for $k$ even, and $(k-1)/2$ for $k$ odd. The highest 
power of $W$ in the numerator is again $W^{2k}$ 
and again only terms with this  highest power (with coefficient 
$B^{k-n}C^{2n}$) will survive when $J_k$ is differentiated $k-$times. 
Accordingly, Eq. (\ref{kyabat}) reduces to    
 \bea                                   
&&S_1(E) =  2\sqrt{2m}\int_{x_{1s}}^{x_{2s}}\sqrt{E-W^2} dx \,\, 
+ \nonumber \\ && \quad 
\hbar \pi\left(1+ \sum_{k=1}^{\infty} 
{(2k-1)!!\over {(k+1)(2k)!!}}
\sum_{n=0}^{n_{max}}{k!\, B^{k-n}C^{2n}\over {(k-2n)!(2n)!}}\right)~.
\label{eq:pot36}
\eea
The main results given earlier by Eqs. (\ref{fine}, \ref{cos}) 
remain valid.    

The summation in eq. (\ref{eq:pot36}) can be done similarly to that in 
Eq. (\ref{eta}). The inner summation provides the mean of 
$(B \pm C \sqrt{B})^k$. Then we find 
 \begin{eqnarray}                       
\eta  &=& \frac{1}{2z_+} [ 1 - \sqrt{1 - z_+} ] + 
\frac{1}{2z_-} [ 1 - \sqrt{1 - z_-} ]\quad {\rm where} \nonumber \\ 
z_\pm &=& B \pm C\sqrt{B} \, . 
\label{sum2}
\end{eqnarray} 
These results (\ref{eta}, \ref{sum2}) are a simple 
demonstration of the relation between WKB and SWKB, which 
Barclay~\cite{barclay} approached in a different manner. 

\begin{figure} 
\scalebox{0.471}{\includegraphics*{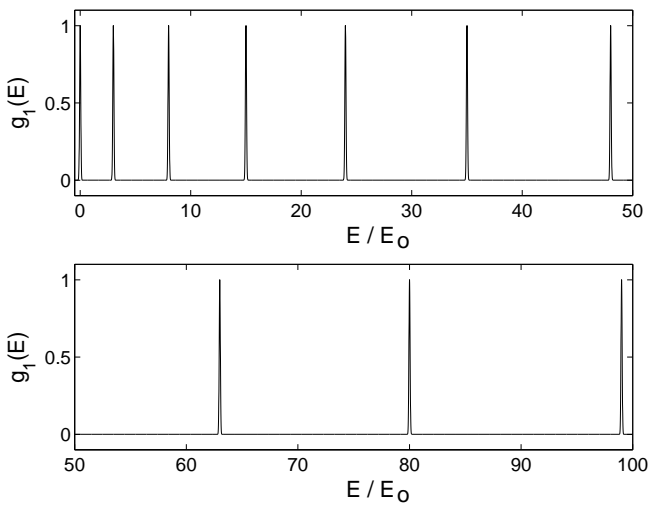}}
\caption{\label{susysqrwell} Numerical evaluation of the trace formula
(\ref{first}) for the infinite square well where
  $F_1(E)=(1+E/E_0)^{1/2}-1$. In the figure, $E$ is plotted in units of 
  $E_0$. To ensure uniform lineshapes, correct
  degeneracies, and strict numerical convergence, we have used the usual
  prescription used in numerical semiclassics (see, for example,
  Section 5.5 of Ref.~\cite{brack}) which is to convolve the
  trace formula with a Gaussian of width $\sigma$. For this particular 
  calculation, we have truncated the sum at $k_{\text{max}}=10^4$ while
  prescribing $\sigma = 0.05$.}
\end{figure}

It may now be instructive to illustrate our results with a few 
examples:\\
1) Infinite Square-well. In this example, $W(x)=-
\hbar\pi/(\sqrt{2m}L) {\rm cot}(\pi x/L)$. It belongs to class I with 
$A=\hbar^2 \pi^2/(2mL^2)=E_0$, $B=1$, and $C=0$. The quantum spectrum 
of $V_1$ is given by $f(n)=n(n+2)E_0$, with $n=0,1,2, \ldots$ . Then 
$F_1(E)=(1+E/E_0)^{1/2}-1$. A careful numerical evaluation of the 
trace formula (\ref{first}) with this $F_1(E)$ reproduces the quantum 
spectrum (see Fig.~\ref{susysqrwell}).  It is also easy to check  
Eq. (\ref{fine}) by 
evaluating the action integral of $W^2(x)$ analytically, and  
Eq. (\ref{semi}) using Eq. (\ref{eta}) ($\eta=1$). 

\noindent 
2) 3-dimensional harmonic oscillator in the $l^{th}$ partial wave. In 
this example $W(r)=\sqrt{2m} \omega r/2-\hbar/(\sqrt{2m}) (l+1)/r$. 
It belongs to class II with $A=\hbar\omega$, $B=1/(2l+2)$ and $C = - 
\sqrt{B}$. The quantum spectrum, measured from the lowest state in a 
fixed partial wave is $f(n)=2n\hbar\omega$, so $F(E) = 
E/(2\hbar\omega)$. Again, Eq. (\ref{fine}) may be checked explicitly. 

To verify Eq. (\ref{semi}), we find from Eq. (\ref{sum2}) that 
 \begin{eqnarray}                       
\eta  &=& \half + \half[ \ell + \half - \sqrt{\ell(\ell+1)} ] 
\end{eqnarray}
in this example. 
The first $1/2$ represents the usual half-integer quantisation in 
LOWKB, while the terms in square brackets arise from the sum of order 
$\hbar^2$ and higher corrections. As discussed in detail by 
Seetharaman \cite{SV84} and Barclay \cite{barclay} they can be 
removed by adopting the Langer prescription. We have also checked 
other examples analytically.  

In conclusion, we have given a new proof that lowest order SWKB 
quantisation is exact, starting from periodic orbit theory, rather 
than by examining the higher order WKB terms. 
The key ingredients have been an invertible algebraic 
expression for the energy spectrum, and Barclay and 
Maxwell's~\cite{barclay, barcmax} insight about shape invariant 
potentials.  

R.K.B and D.W.L.S are grateful to NSERC for continuing research
support under discovery grants. 



\begin{thebibliography}{99}

\bibitem{cooper1}
F. Cooper, A. Khare, and U. Sukhatme, {\it Supersymmetry in Quantum 
Mechanics}, (World Scientific, Singapore, 2001), p. 36.  

\bibitem{migdal} A. B. Migdal and V.P. Krainov, {\it Approximate Methods in 
Quantum Mechanics} (W. A. Benjamin Inc., N.Y., 1969) p. 119.

\bibitem{langer} 
R. E. Langer, Phys. Rev. {\bf 51}, 669 (1937).
\bibitem{barclay} 
D. T. Barclay, Phys. Lett. {\bf A 185}, 169 (1994). 

\bibitem{gend} 
L. Gendenshtein, JETP Lett. {\bf 38}, 356 (1983).

\bibitem{comtet} A. Comtet, A.D. Bandrauk and D.K. Campbell, 
Phys. Lett. {\bf B 150}, 159 (1985). 
\bibitem{khare}
R. Dutt, A. Khare and U. Sukhatme, Phys. Lett. {\bf B 181}, 295 (1986); 
R. Adhikari, R. Dutt, A. Khare and U. Sukhatme, Phys. Rev. {\bf A 38}, 
1679 (1988).

\bibitem{brack}
M. Brack and R. K. Bhaduri, {\it Semiclassical Physics} (Westview Press,
Boulder, Colorado, 2003).

\bibitem{dutt}
R. Dutt, A. Khare, and U. Sukhatme, Am. J. Phys. {\bf 56}, 163 (1988).

\bibitem{Balian}
R. Balian and C. Bloch, Ann. Phys. (N. Y.) {\bf 60}, 401 (1970); 
{\bf 63}, 592 (1971); {\bf 69}, 76 (1972).

\bibitem{jennings}
B. K. Jennings, Ph.D. Thesis, McMaster University, 1976 (unpublished);

\bibitem{ross}
R. K. Bhaduri and C. K. Ross, Phys. Rev. Lett. {\bf 27}, 606 (1971);
B. K. Jennings, Ann. Phys. (N. Y.) {\bf 84}, 1 (1974).

\bibitem{barcmax} D. T. Barclay and C J. Maxwell, Phys. Lett. 
{\bf A 157}, 357 (1991). 

\bibitem{SV84}
M. Seetharaman and S.S. Vasan, J. Phys. A {\bf 17}, 2485 (1984).

\end{thebibliography}
\end{document}